\documentclass{llncs}
\usepackage[utf8]{inputenc}

\usepackage{stmaryrd}
\usepackage{bussproofs}
\usepackage{multirow}

\usepackage{mathptmx}

\usepackage{latexsym}
\usepackage{amsfonts}
\usepackage{amssymb}
\usepackage{amsmath}
\usepackage{graphicx}
\usepackage{color}

\usepackage{wrapfig,lipsum,booktabs}
\usepackage{semantic}
\usepackage{listings}
\usepackage{proof}

\usepackage{subcaption}
\usepackage{url}

\usepackage{tikz}
\usepackage{mathdots}
\usepackage{yhmath}
\usepackage{cancel}
\usepackage{siunitx}
\usepackage{array}
\usepackage{gensymb}
\usepackage{tabularx}
\usepackage{booktabs}
\usetikzlibrary{fadings}

\usepackage{pifont}
\usepackage{floatrow}
\newfloatcommand{capbtabbox}{table}[][\FBwidth]

\makeatletter
\newcommand\tabcaption{\def\@captype{table}\caption}
\newcommand\figcaption{\def\@captype{figure}\caption}
\makeatother

\definecolor{codegreen}{rgb}{0,0.6,0}
\definecolor{codegray}{rgb}{0.5,0.5,0.5}
\definecolor{codepurple}{rgb}{0.58,0,0.82}
\definecolor{backcolour}{rgb}{0.95,0.95,0.92}

\lstdefinestyle{csp}{
    backgroundcolor=\color{backcolour},
    commentstyle=\color{codegreen},
    keywordstyle=\color{magenta},
    numberstyle=\tiny\color{codegray},
    stringstyle=\color{codepurple},
    basicstyle=\footnotesize,
    breakatwhitespace=false,
    breaklines=true,
    captionpos=b,
    keepspaces=true,
    numbers=left,
    numbersep=5pt,
    showspaces=false,
    showstringspaces=false,
    showtabs=false,
    tabsize=4,
    morekeywords={var,define,Stop,Skip,if,else,assert}
}

\usepackage{tikz}
\usepackage{standalone}

\usepackage{algorithm}
\usepackage{algcompatible}

\usepackage[noend]{algpseudocode}

\algblockdefx{ForAllP}{EndFAP}[1]%
  {\textbf{for all }#1 \textbf{do in parallel}}%
  {\textbf{end for}}

\begin{document}


%
\author{Hadrien Bride\inst{1}, Cheng-Hao Cai\inst{2}, Jin Song Dong\inst{1,3}, Rajeev Gore\inst{4}, Zh\'e H\'ou\inst{1}, Brendan Mahony\inst{5} and Jim McCarthy\inst{5}}
\authorrunning{Bride et al.} 
%
\institute{
Institute for Integrated and Intelligent Systems, Griffith University, Australia
  \and
  School of Computer Science, University of Auckland, New Zealand
  \and
  School of Computing, National University of Singapore, Singapore
  \and
  Research School of Computer Science, The Australian National University, Australia
  \and
  Defence Science and Technology, Australia
  }

\title{N-PAT: A Nested Model-Checker}
\subtitle{(System Description)}
\maketitle

\begin{abstract}

N-PAT is a new model-checking tool that supports the verification of nested-models, i.e. models whose behaviour depends on the results of verification tasks. In this paper, we describe its operation and discuss mechanisms that are tailored to the efficient verification of nested-models. Further, we motivate the advantages of N-PAT over traditional model-checking tools through a network security case study.

\end{abstract}

\section{Introduction}
\label{sec:intro}

Model-checking is the problem of formally verifying that a model of a system meets a given specification. 
Automated model-checking techniques have been successfully applied to find subtle errors in complex industrial designs of e.g., hardware circuits, software controllers, and communication protocols~\cite{clarke2018}.
However, the adoption rate of model-checking
remains low in software engineering because of the computational complexity of model-checking algorithms. The state-space explosion problem~\cite{clarke2011model} makes the verification of large models intractable unless high-level abstractions are used in the development and leveraged during verification.

Nowadays, complex systems are often designed in a modular and hierarchical fashion. Hierarchical models, also called multilevel models, are abstract representations of systems that span multiple levels of abstraction. They encode the hierarchical structure of systems explicitly and therefore enable reasoning about how properties of one level reflect across multiple levels of the model~\cite{parvu2016novel}.

In this paper, we introduce the notion of nested model and nested model-checking. The main idea is to break up a large model-checking task into a hierarchy of smaller model-checking tasks. 
A Nested model is a high-level model which may contain several child models nested inside; its behaviour depends on the verification results of its child models. Note that the properties to be verified in child tasks may be different from the properties to be verified in  parent tasks.

We present N-PAT -- a nested model-checker suited to the verification of hierarchical systems and designed to perform nested model-checking tasks.
In hierarchical modelling, some verification tasks may be used to determine and lift the properties of underlying child models to parent models. This structural abstraction provides modellers with the ability to structure the verification and guide the state-space exploration of model-checking methods, it also provides significant benefits in term of scalability for verification when compared to the traditional approach to modelling. We implement several optimisations leveraging the hierarchical structure of nested models. Also, since the time and space complexity of model-checking algorithms with respect to the size of models is super-linear, the divide-and-conquer approach employed by N-PAT significantly reduces the overall verification time.
What sets N-PAT apart from existing model checkers (e.g., \cite{sun2008model}, \cite{steffen2018m3c}, \cite{lopez2014meta}) is the abstraction level of the modelling language.
In our work, the modelling language of nested models has high-level primitives such as model checking and nested model instantiation. 
\section{Nested Model-checking}
\label{sec:prelim}

\textit{Standard} model checking is the problem of verifying whether a \textit{standard} model complies with a given property. A standard model is a static and finite-state representation of a system, which may exhibit non-deterministic and probabilistic behaviours. The semantics of a standard model can be specified as a labelled transition system or a Markov decision process. Properties that can be verified include reachability, deadlock-freeness, divergence-freeness, reachability, and LTL formulae. The result of a model checking task depends on the type of the model. When checking a property over a non-probabilistic model, the result is $0$ (not satisfied) or $1$ (satisfied). When checking a property over a probabilistic model, the result is the min (alt. max) probability that the property is satisfied. Note that we only consider results in natural numbers, and a probability is represented in e.g., per thousand. Formally, let $M^s$ be the set of standard models and $\Phi$ be the set of properties. We denote by $\texttt{mc} : M^s \times \Phi \to \mathbb{N}$ the model checking function that returns the results of checking a property over a standard model.

A \textit{meta model}, also commonly referred to as a \emph{template}, is a model of standard models. It can be viewed as a function that has a finite number of arguments and returns a standard model. Formally, a meta model is a function of the form $A_1 \times ... \times A_n \to M^s$ where $n \in \mathbb{N}$ and $A_1,...,A_n \in \mathbb{N}$. We denote by $M^m$ the set of meta models. 
In order to instantiate a meta model, every argument must be known. An instantiated meta model is a standard model and can be verified using standard model checking.

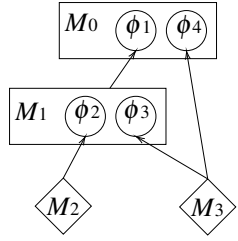
\begin{figure}[h!]
    \centering
    \tikzset{every picture/.style={line width=0.75pt}} 

\begin{tikzpicture}[x=0.75pt,y=0.75pt,yscale=-1,xscale=1]

\draw   (230,216) .. controls (230,202.19) and (241.19,191) .. (255,191) .. controls (268.81,191) and (280,202.19) .. (280,216) .. controls (280,229.81) and (268.81,241) .. (255,241) .. controls (241.19,241) and (230,229.81) .. (230,216) -- cycle ;
\draw   (160.5,181) -- (355.5,181) -- (355.5,251) -- (160.5,251) -- cycle ;
\draw   (293,216) .. controls (293,202.19) and (304.19,191) .. (318,191) .. controls (331.81,191) and (343,202.19) .. (343,216) .. controls (343,229.81) and (331.81,241) .. (318,241) .. controls (304.19,241) and (293,229.81) .. (293,216) -- cycle ;
\draw   (293,109) .. controls (293,95.19) and (304.19,84) .. (318,84) .. controls (331.81,84) and (343,95.19) .. (343,109) .. controls (343,122.81) and (331.81,134) .. (318,134) .. controls (304.19,134) and (293,122.81) .. (293,109) -- cycle ;
\draw   (222.5,74) -- (418.5,74) -- (418.5,144) -- (222.5,144) -- cycle ;
\draw   (356,109) .. controls (356,95.19) and (367.19,84) .. (381,84) .. controls (394.81,84) and (406,95.19) .. (406,109) .. controls (406,122.81) and (394.81,134) .. (381,134) .. controls (367.19,134) and (356,122.81) .. (356,109) -- cycle ;
\draw    (285.5,181) -- (316.86,135.65) ;
\draw [shift={(318,134)}, rotate = 484.66] [color={rgb, 255:red, 0; green, 0; blue, 0 }  ][line width=0.75]    (10.93,-3.29) .. controls (6.95,-1.4) and (3.31,-0.3) .. (0,0) .. controls (3.31,0.3) and (6.95,1.4) .. (10.93,3.29)   ;

\draw    (227.5,293) -- (254.07,242.77) ;
\draw [shift={(255,241)}, rotate = 477.87] [color={rgb, 255:red, 0; green, 0; blue, 0 }  ][line width=0.75]    (10.93,-3.29) .. controls (6.95,-1.4) and (3.31,-0.3) .. (0,0) .. controls (3.31,0.3) and (6.95,1.4) .. (10.93,3.29)   ;

\draw    (407.5,294) -- (319.72,242.02) ;
\draw [shift={(318,241)}, rotate = 390.63] [color={rgb, 255:red, 0; green, 0; blue, 0 }  ][line width=0.75]    (10.93,-3.29) .. controls (6.95,-1.4) and (3.31,-0.3) .. (0,0) .. controls (3.31,0.3) and (6.95,1.4) .. (10.93,3.29)   ;

\draw    (407.5,294) -- (381.33,135.97) ;
\draw [shift={(381,134)}, rotate = 440.6] [color={rgb, 255:red, 0; green, 0; blue, 0 }  ][line width=0.75]    (10.93,-3.29) .. controls (6.95,-1.4) and (3.31,-0.3) .. (0,0) .. controls (3.31,0.3) and (6.95,1.4) .. (10.93,3.29)   ;

\draw   (227.5,293) -- (262.5,328) -- (227.5,363) -- (192.5,328) -- cycle ;
\draw   (407.5,294) -- (442.5,329) -- (407.5,364) -- (372.5,329) -- cycle ;

\draw (250,99) node   [align=left] {{\Huge $M$}{\LARGE $0$}};
\draw (188,210) node   [align=left] {{\Huge $M$}{\LARGE $1$}};
\draw (227.5,328) node   [align=left] {{\Huge $M$}{\LARGE $2$}};
\draw (407.75,330) node   [align=left] {{\Huge $M$}{\LARGE $3$}};
\draw (321,104) node   [align=left] {{\Huge $\phi$}{\LARGE $1$}};
\draw (257,211) node   [align=left] {{\Huge $\phi$}{\LARGE $2$}};
\draw (319,211) node   [align=left] {{\Huge $\phi$}{\LARGE $3$}};
\draw (384,104) node   [align=left] {{\Huge $\phi$}{\LARGE $4$}};

\end{tikzpicture}
    \caption{Illustration of the structure of a nested model.}
    \label{fig:tier1}
\end{figure}

Traditionally, meta models are instantiated from values specified by the modeller. In our work, we consider the verification of meta models instantiated from values that are the result of model checking tasks. Such a meta model is called a \textit{nested} model and denoted as $M^n$.
Figure~\ref{fig:tier1} illustrates the structure and components of a nested model. Each diamond represents a standard model. Each box represents a meta model. Verification tasks are symbolised by circles. The text within each circle is the property to be checked in the corresponding verification task. The arrows symbolise dependencies among verification tasks. In Figure~\ref{fig:tier1}, there are two standard models: $M_2$ and $M_3$, and two meta models: $M_0$ and $M_1$. For instance, $M_1$ requires the verification results from $\texttt{mc}(M_2, \phi_2)$ and $\texttt{mc}(M_3, \phi_3)$ to be instantiated. After instantiation, $M_1$ will become a standard model. To verify the property $\phi_0$ over the nested-model $M_0$, we need to evaluate the following expression:
$\texttt{mc}(M_0(\texttt{mc}(M_1(\texttt{mc}(M_2, \phi_2), \texttt{mc}(M_3, \phi_3)), \phi_1), \texttt{mc}(M_3, \phi_4)), \phi_0)$.

A nested model checking problem is an expression that can be evaluated to an integer; it is formulated in a language that has two main primitives: meta model instantiation and standard model checking. For convenience, the language of nested model checking problems is extended with integers, basic arithmetic operations and restricted scope constant definitions. Let $const$ be the set of constant identifiers, the syntax of nested model checking problems is given by the grammar in Figure~\ref{fig:bnf-nm-cd}, where $[\alpha]^{*}$ denotes repeating $\alpha$ zero or more times. 

\vspace{-10px}
\begin{figure}[h!]

\begin{tabular}{l@{\hskip 20px}l}
    $binding ::= const~ \textbf{=} ~expr$ &  $mc ::= \textbf{mc(} model \textbf{,}~ \Phi \textbf{)}$\\
    $model ::= M^s \mid M^m \textbf{(} binding~ [\textbf{,}~ binding]^{*} \textbf{)}$ & $op ::= expr~ (+ \mid - \mid \textbf{$\times$} \mid /) ~expr$\\
    $def ::= \textbf{let}~ binding~ [\textbf{,}~ binding]^{*} ~\textbf{in}~ expr$ & $expr ::= \mathbb{N} \mid op \mid mc \mid const \mid def$
    
\end{tabular}
\caption{The BNF grammar of nested model checking problems.}
\label{fig:bnf-nm-cd}
\end{figure}
\vspace{-10px}

We assume that the set of verification tasks related to a nested model form a directed acyclic graph (as in Figure~\ref{fig:tier1}), which defines the dependencies of tasks, and that the tasks and the graph are known before the verification stage.
When a \texttt{let} expression introduces multiple bindings, these bindings must be independent of one another, non-recursive, and may be evaluated in parallel. 

The semantics of nested model checking problems in a given \textit{context} is defined by the \textit{evaluation function} \texttt{eval}, given in Figure~\ref{fig:sem-nm-cd}. A \textit{valued} binding is a tuple $\langle c,v\rangle $ where $c \in \mathit{const}$ and $v \in \mathbb{N}$. A context is a set of valued bindings. Let $\Gamma$ be a context,  $e, e_1,... \in \mathit{expr}$ be expressions, $c, c_1,... \in \mathit{const}$ be constant identifiers, $m_s \in M^s$ be a standard model, $m_m \in M^m$ be a meta model, $m \in M^s \cup M^m$ be a model, $\phi \in \Phi$ be a property, $n, v, v_1, ... \in \mathbb{N}$ be numbers, and $op \in \{+,-,*,/\}$ be an integer operator. 

\vspace{-10px}
\begin{figure}[h!]

\begin{tabular}{l@{\hskip 49px}l@{\hskip 48px}l}
eval($n$, $\Gamma$) = $n$ & eval($m_s$, $\Gamma$) = $m_s$ & eval($c$, $\Gamma$) = $v$ where $\langle c,v\rangle  \in \Gamma$ \\[5px]
\multicolumn{3}{l}{eval($e_1~~op~~e_2$, $\Gamma$) = eval($e_1$, $\Gamma$) $~op~$ eval($e_2$, $\Gamma$) \hspace{30px} eval(mc($m$, $\phi$), $\Gamma$) = mc(eval($m$, $\Gamma$), $\phi$)}\\[5px]
\multicolumn{3}{l}{eval($m_m$($\{\langle c_1,e_1\rangle ,...,\langle c_n,e_n\rangle \}$), $\Gamma$) = m($\{\langle c_1,v_1\rangle ,...,\langle c_n,v_n\rangle \}$)
    where }\\
    \multicolumn{3}{l}{\hspace{20px} $v_1=~$eval($e_1$,  $\Gamma$), ..., $v_n=~$eval($e_n$, $\Gamma$)}\\[5px]
\multicolumn{3}{l}{eval(let $\{\langle c_1,e_1\rangle ,..., \langle c_n,e_n\rangle \}$ in $e$) = eval($e$, $\Gamma'$) 
    where }\\
    \multicolumn{3}{l}{\hspace{20px} $\Gamma' = \Gamma \cup \{\langle c_1,$eval($e_1$, $\Gamma$)$\rangle ,..., \langle c_n,$eval($e_n$, $\Gamma$)$\rangle \}$}
\end{tabular}
\caption{The semantics of nested model checking problems.}
\label{fig:sem-nm-cd}
\end{figure}

\section{N-PAT: Implementation}
\label{sec:implem}

N-PAT is built on top of Process Analysis Toolkit~\cite{sun2008model} (PAT) -- an industrial scale model-checker which employs an expressive modelling language called Communicating Sequential Processes with C\# (CSP\#) developed by Hoare~\cite{Hoare1978} and others~\cite{sun2009}. PAT features a model editor and an animated simulator using a mature and IDE-style user interface. 
Further, PAT facilitates new language and algorithm design and implementation as extended modules. Over the past 10 years, we have extended PAT with new verification modules for timed automata~\cite{sun2009}, real-time systems~\cite{Sun2013}, and probabilistic systems~\cite{Sun2010}. 
We implemented N-PAT in C\# as an extension of PAT. N-PAT is open-source and freely available online.\footnote{\url{https://formal-analysis.com/research/npat/index.html}}

Standard models are specified by (probabilistic) CSP\# models and properties are specified by CSP\# assertions~\cite{sun2008model}. Meta models are specified by a \textit{meta-level} CSP\# language. This language, called meta-CSP\#, introduces labelled place-holders, of the form $[\mathit{id}]$ where $\mathit{id} \in \mathit{const}$ is a label. These labelled place-holders extend the CSP\# language and can be used in place of integer constants (e.g., variable initial values, choice probabilities). Let $m$ be a meta-CSP\# model and $id_1,...,id_n$ where $n \in \mathbb{N}$ be the set of placeholders that appears in its definition. Let $v = \{\langle id_1,v_1\rangle ,...,\langle id_n,v_n\rangle \}$ where $v_1, ..., v_n \in \mathbb{N}$ be a set of \textit{valued} bindings.  The meta-CSP\# model $m$ can be instantiated using $v$ into a CSP\# model by substituting the occurrences of $[\mathit{id}_i]$ by $v_i$ for all $i \in \{1,...,n\}$. Nested model checking problems are specified using the language described in Section~\ref{sec:prelim}. Model checking is performed by N-PAT through the orchestration of calls to PAT.

\begin{figure}[h!]
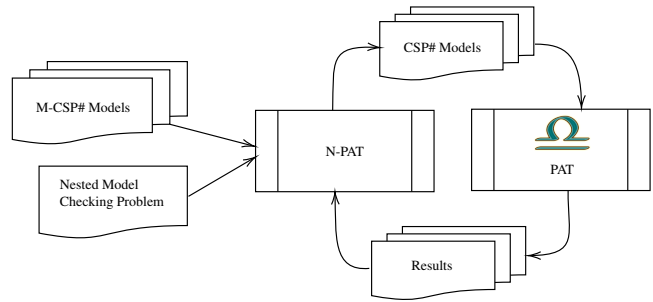

    \centering
    \includestandalone[width=0.7\textwidth]{img/arch}
    \caption{N-PAT data-flow overview.}
    \label{fig:arch}
\end{figure}

\vspace{-0.5cm}

Figure~\ref{fig:arch} depicts the overall data-flow of N-PAT.  The input of N-PAT is a set of standard CSP\# and meta-CSP\# models and a nested model checking problem. N-PAT evaluates the result of the nested model checking problem similarly to how a dynamic interpreter evaluates an expression. The nested model checking problem is first parsed, and a corresponding abstract syntactic tree is built. This step is implemented using parser combinators (i.e., using recursive descent parsing). The resulting abstract syntactic tree is then recursively evaluated in a bottom-up fashion. 

N-PAT exploits the hierarchical nature of nested models and provides improved verification scalability when compared to traditional model checkers that operate on flattened models. First, since CSP\# models are static (i.e., they are not modified during execution), N-PAT applies stage-wise partial evaluation of verification sub-tasks to optimise the verification phase of nested CSP\# models.
Second, given a nested CSP\# model, we assume that its verification sub-tasks are independent and can be computed concurrently. Thus N-PAT uses parallelism to speed up verification on modern architectures with multiple cores. This parallelism manifests itself in three places: bindings, operands, and 
the evaluation of meta model arguments.

\section{Case Study and Experiment}
\label{sec:exp}

We introduce a network security case study to illustrate the modelling and scalability advantages of nested model-checking. The case study is concerned with computing the probabilistic security level of a network. It is a simplified version of a real-life example which is studied by Australian Defence. The problem is hierarchical by nature, which illustrates code-reuse and modularisation of the proposed modelling approach. Another nice property of this example is that we can create models of different sizes to test the scalability of the verification, as will be shown in the experiment.

The details of the example are as follows: suppose there is a cluster of computation nodes, and each node can be in one of the three states: \emph{safe}, \emph{compromised}, and \emph{isolated}. Initially, each node is safe, but it has a chance to be \emph{hacked}, which changes the state of the node to compromised. When a node is compromised, it can either be \emph{patched}, which will make the node safe again, or be \emph{isolated}, which will disconnect the node from the cluster. If a node is isolated, then it loses the connection to other nodes and thus cannot contribute to the computational power of the cluster. When the node is isolated, it has a chance to be \emph{recovered}, which will lead the node to the safe state again. Otherwise, the node stays isolated, in which case we increase \texttt{down\_nodes\_counter}, i.e., the number of nodes that are offline, by 1. For simplicity, we assume that the hacking of each node is independent. We model two types of nodes: \emph{normal} node and \emph{premium} node, where the latter has a higher chance to be patched when it is compromised and a higher chance to be recovered when it is isolated.

\begin{figure}[h!]
  \centering
  \includegraphics[width = \textwidth]{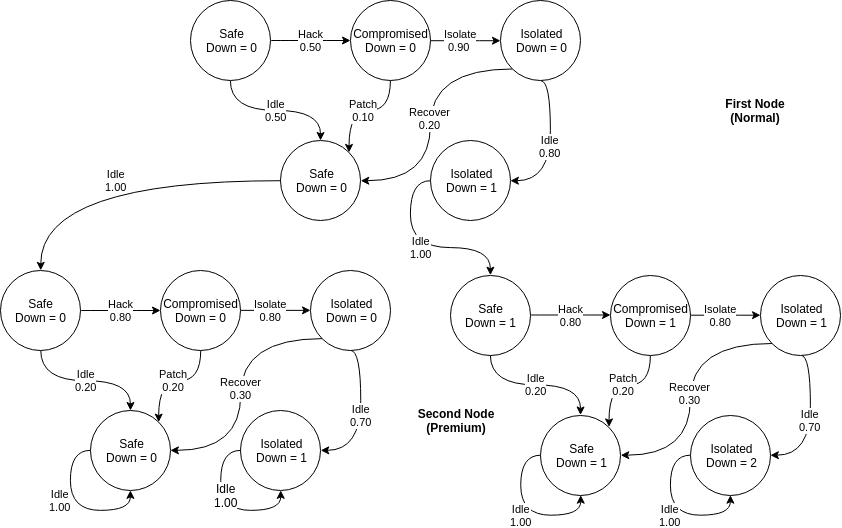}
  \caption{The Markov chain of the traditional model, exemplified with two nodes.}
  \label{fig:flattened}
\end{figure}

\paragraph{Traditional Method:} 
We shall formalise the above as a Markov chain in CSP\#, and we have to define the state transitions for both types of nodes. Next, we model a \texttt{cluster manager}, which iterates through each node in the cluster and checks whether the node is offline. The manager will report that the cluster is in critical condition if at least half of the nodes in the cluster are offline. Consider an example where there are only two nodes in the network: the first node is a normal node, and the second node is a premium node. We show the overall Markov chain in Figure~\ref{fig:flattened}. We annotate the event and its probability on the arrows. The upper part of the figure describes the process of checking the normal node, which can be either safe or isolated.
The lower part of the figure splits into two cases when checking the premium node.
The \texttt{Down = x} line in each circle indicates the \texttt{down\_nodes\_counter}. The premium node in Figure~\ref{fig:flattened} has a higher chance to be hacked than because in our hypothetical scenario, the hacker is more likely to target a premium node.

\begin{figure}[t!]
    \centering
    \begin{subfigure}[b]{0.49\textwidth}
        \includegraphics[width=\textwidth]{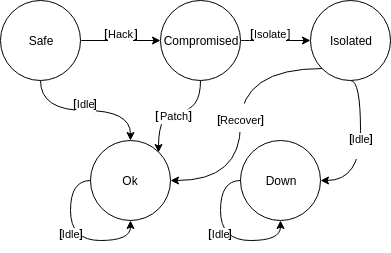}
        \caption{Node Level}
        \label{fig:node_level}
    \end{subfigure}
    ~ 
    \begin{subfigure}[b]{0.49\textwidth}
        \includegraphics[width=\textwidth]{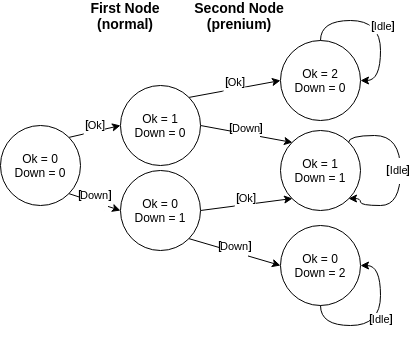}
        \caption{Cluster Level}
        \label{fig:cluster}
    \end{subfigure}
    \caption{Modular models of Figure~\ref{fig:flattened} for nested model checking.}\label{fig:MMCModels}
\end{figure}

\paragraph{Nested Model Checking Method:} 
We break the model in Figure~\ref{fig:flattened} down into two levels of abstraction: the node level and the cluster level. The idea is to create more modular models so that the model for a node can be used for both normal nodes and for premium nodes, and potentially can be used in future developments of other models. 
At the node level, we study the common properties of the two types of nodes, and try to generalise the model so that the model-checking result is exactly what we need at the cluster level. As the cluster manager, we only need to know whether a node is offline or not. Therefore, besides safe, isolated, and compromised, we give a node two additional states: \texttt{ok} and \texttt{down}. Like in the traditional model, each node is initialised to be in the safe state. 
Since we do not consider cluster manager operations at the node level, we do not use the counter for offline nodes. Instead, when the
node stays isolated, we change its state to \texttt{down}, and when it is not hacked/patched/recovered, we change its state to \texttt{ok}. The state transition diagram is shown in Figure~\ref{fig:node_level}.

At the cluster level, we abstract the notion of a node into only two states: \texttt{ok} and \texttt{down}. The probability of going to these two states will be given by the model-checking results from the node level. If a node is down, then we increment \texttt{down\_nodes\_counter}. We then model the cluster manager similarly as in the traditional model. The (partial) Markov chain for checking nodes is illustrated in Figure~\ref{fig:cluster}, where we only show two nodes. 
Note that places holders for probabilities in Figure~\ref{fig:node_level}, such as \texttt{[Isolated]},  will be instantiated with specific values, depending on the type of the node, at run-time. Place holders in Figure~\ref{fig:cluster} will be instantiated with the results of node-level verification at run-time.
Compared to the traditional model (cf.~Figure~\ref{fig:flattened}), the nested model is much simpler, and we will show that this leads to significant improvement in
scalability.

\paragraph{Experimental comparison:} We compare the performance of both modelling approaches by evaluating the overall security of network of different sizes. In each case, the number of normal nodes are $4/5$ of the total number. All experiments were carried on a desktop with Core i7-7700 quad-core processor at 3.6GHz and 32GB RAM. 
We verify multiple instances of the above models, starting with 8 nodes in a cluster, and increase the number of nodes by 2 at a time, and maintain the number of normal nodes at $(4 \times num\_of\_nodes) / 5$. 
We then observe the time used in model checking as the number of nodes increases. We run each test 5 times and compute the average time spent to obtain the results.

\vspace{-10px}
\begin{table}[h!]
\small
\begin{tabular}{|l|c|c|c|c|c|c|c|c|c|c|c|c|c|c|}
  \hline
  \textbf{Number of Nodes} & \textbf{8} & \textbf{10} & \textbf{12} & \textbf{14} & \textbf{16} & \textbf{18} & \textbf{20} & \textbf{22} & \textbf{24} & \textbf{26} & \textbf{28} & \textbf{30} & \textbf{32} & \textbf{34}\\
  \hline
Runtime Traditional (ms) & 248 & 306 & 569 & 1771 & 7411 & 35K & 279K & \multicolumn{7}{c|}{Out of memory}\\
  \hline
  Runtime N-PAT (ms) & 427 & 430 & 430 & 430 & 430 & 431 & 445 & 438 & 442 & 461 & 458 & 469 & 465 & 476\\
  \hline
\end{tabular}
\caption{Experiment of traditional (probabilistic) model-checking compared with nested model checking.}
\label{tab:exp}
\end{table}
\vspace{-10px}

\paragraph{Discussion:} As seen from the results in Table~\ref{tab:exp}, the run-time of traditional model-checking grows rapidly as the size of the model (the number of nodes) increases. On the other hand, the run-time growth of nested model checking is moderate, and it solves instances up to 34 nodes less than 0.5 seconds. N-PAT also uses very little memory compared to PAT which uses up to 26.6GB memory when running the 20 nodes instance. For small examples, the traditional modelling approach may be faster because the verification of nested models involves several calls to PAT which incur marginal overhead. However, the nested model checking approach scales better. The source code of this experiment, i.e. both the traditional model and the hierarchical model,  
can be found online.\footnote{\url{https://formal-analysis.com/research/npat/examples.html}}

\section{Conclusion and Future Work}
\label{sec:future}

We presented N-PAT -- a high-level model checker that enables the verification of models that relies on the results of other verification tasks. We demonstrated in a case study in network security that this tool permits the use of high-level abstraction mechanisms and can therefore significantly improve the time and memory efficiency of verification tasks. These results indicate that nested model checking provides a novel modelling approach that can in some cases scale better than traditional model-checking. 

In future work, we intend to provide more modelling flexibility by allowing dynamic calls to verification tasks that are not known a priori. We also planned to apply dynamic language optimisation techniques such as memoisation to speed up verification. Finally, we planned on supporting a fully reflective modelling language that permits inspection and modification of the behaviour and structure of models at verification-time.


\bibliographystyle{plain}
\bibliography{main}


\end{document}